# Patched MOA: optimizing inference for diverse software development tasks

Asankhaya Sharma, Patched Codes, Inc
asankhaya@patchedcodes.com

## Abstract

This paper introduces Patched MOA (Mixture of Agents), an inference optimization technique that significantly enhances the performance of large language models (LLMs) across diverse software development tasks. We evaluate three inference optimization algorithms—Best of N, Mixture of Agents, and Monte Carlo Tree Search—and demonstrate that Patched MOA can boost the performance of smaller models to surpass that of larger, more expensive models. Notably, our approach improves the gpt-4o-mini model's performance on the Arena-Hard-Auto benchmark by 15.52%, outperforming gpt-4-turbo at a fraction of the cost. We also apply Patched MOA to various software development workflows, showing consistent improvements in task completion rates. Our method is model-agnostic, transparent to end-users, and can be easily integrated into existing LLM pipelines. This work contributes to the growing field of LLM optimization, offering a cost-effective solution for enhancing model performance without the need for fine-tuning or larger models. Our implementation is open-source and available at https://github.com/codelion/optillm.

# Introduction

In the past year, the typical LLM inference workload has steadily moved away from single query-response per request to complex multi-step reasoning workflows. Agentic workflows that make multiple calls to LLM require inference to be fast, cheap and accurate. This presents opportunities for several trade-offs when it comes to choosing the model that is most appropriate for a given task. In addition, there has been work to see if smaller or less capable models can be used to do the same task with same performance as the bigger and more capable model by guiding the inference. In this article, we evaluate three different approaches that can be applied during inference to improve the performance of the models on underlying tasks.

Our approach is generic, works with any kind of model or downstream task and is completely transparent to the end user. We show that, in general, optimizing inference by calling the same model with guided prompts via multiple API calls improves the overall performance.

In particular, our key contributions are:

- We benchmark and evaluate three inference optimization algorithms from the literature (best of n, mixture of agents and monte carlo tree search). Our implementation is open-source and is included in optillm for anyone to try out.
- We find that all these approaches present different trade-offs in terms of speed, cost and accuracy.
- Patched MOA boosts gpt-4o-mini (by 15.52%) to the top of Arena-Hard-Auto benchmark with only a fraction of the cost it takes to run gpt-4-turbo (the current best model).

# Approach

We evaluated the following three techniques.

1) **Best of N (bon):**

   We use the query and generate 3 (n=3) responses (R1, R2, R3) from the model [1], we then score (N1, N2, N3)) the responses and pick the one that is the best. The scoring can be done using another reward model or a metric or in our case as we are interested in optimizing inference using only a single LLM we use the same LLM to generate the scores.

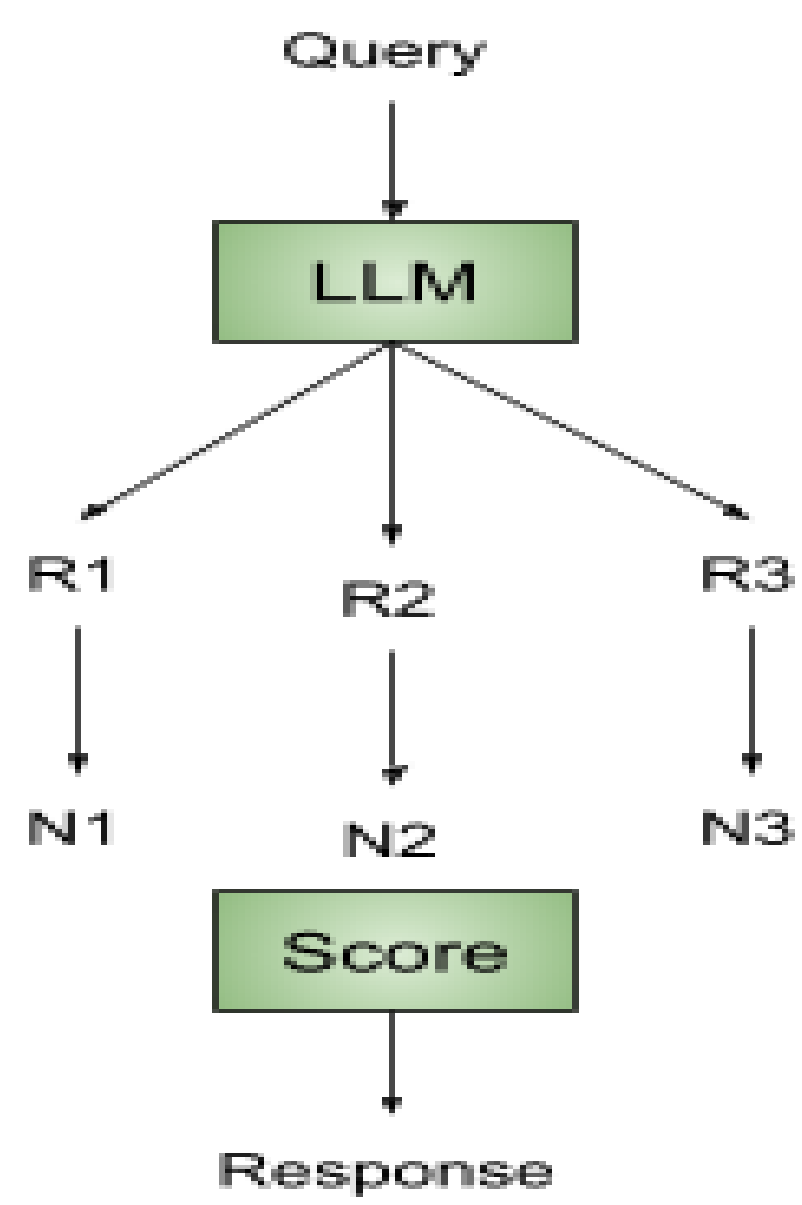

2) **Mixture of Agents (moa):**

   Recently, mixture of agents approach introduced by Together AI [2] has shown promise and was found to outperform GPT-4. In this approach, the query is first used to generate 3 (n=3) responses (R1, R2, R3) and then the model is used to generate 3 corresponding critiques (C1, C2, C3) of the responses. Finally, the model is given the original query, initial responses and the critiques to generate a final response.

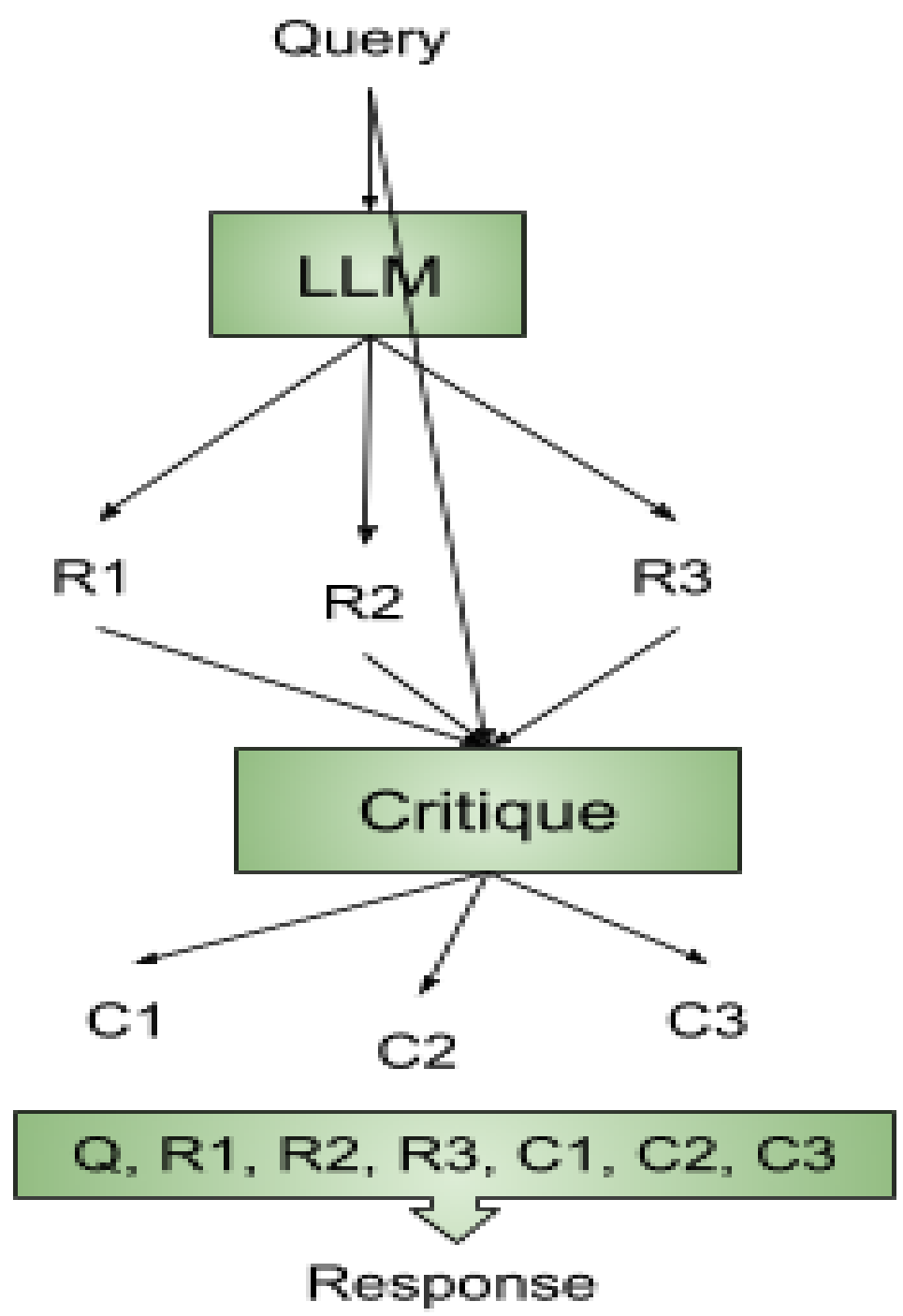

3) **Monte Carlo Tree Search (mcts):**

Monte Carlo Tree Search can be used to explore different dialogue states and generate responses from the same LLM [3] to find high quality responses that lead to good outcomes. In our experiments with MCTS we set depth = 1 and simulation = 2 and exploration = 0.2. For each simulation, we initially start by generating 3 responses (R1, R2, R3) from the original query. We then prompt the model to generate another set of corresponding queries (Q1, Q2, Q3) to further clarify the initial query or explain the initial response. We then expand the tree with these queries to generate the set of responses (R4, R5, R6).The conversations are then evaluated (we use the same model to generate a score to evaluate the quality of the conversation) and back propagated to the root. Finally, the model with the highest UCB1 score is selected.

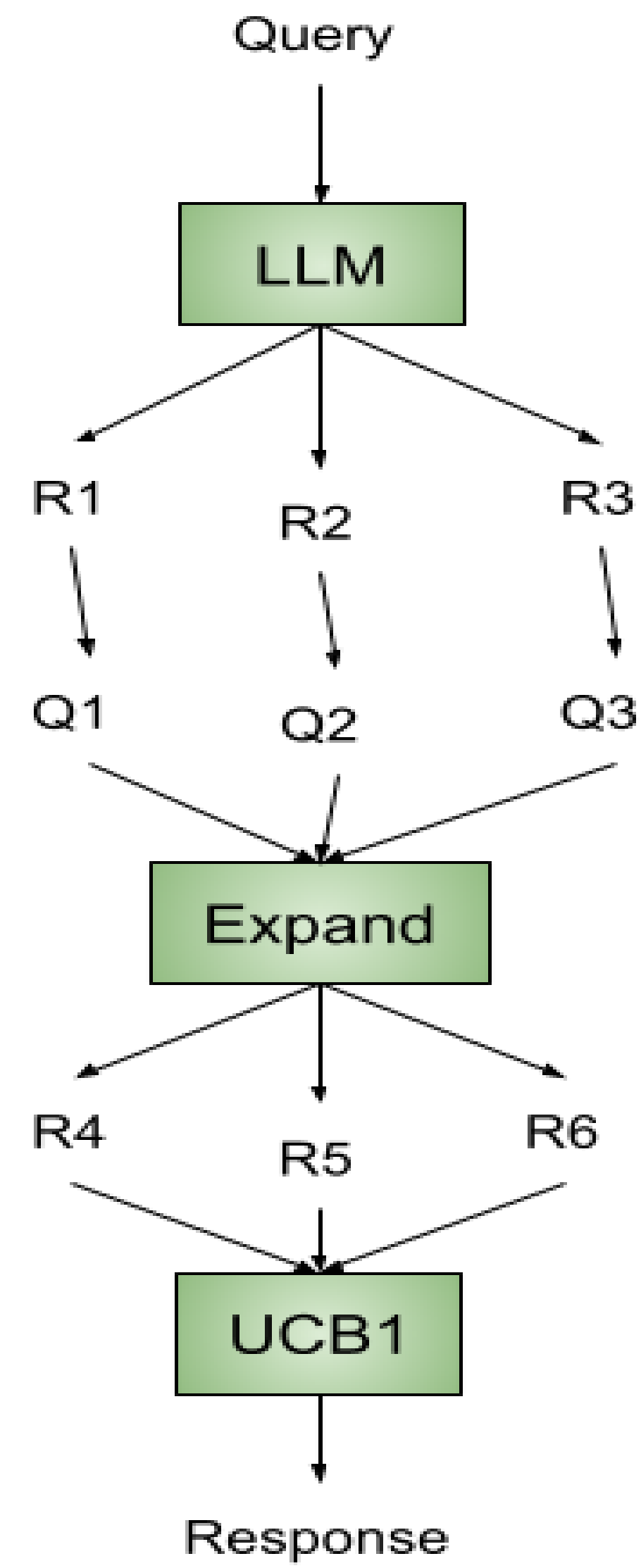

These three different optimization techniques represent different trade-offs in terms of time and cost. Assuming multiple responses generated from the model are of similar length and certain calls can be done in parallel, we estimate the cost and time below.

| Optimization | Calls | Time | Cost |
|---|---|---|---|
| bon | 4x | 3x | 4x |
| moa | 3x | 8x | 8x |
| mcts | 9x | 8x | 32x |

As is clear the most expensive optimization is the mcts due to the higher number of calls and cost. While bon is the most cheaper optimization. Next, we will evaluate the performance of these techniques to gauge if they indeed improve the accuracy over a diverse range of tasks.

# Evaluation

In order to evaluate the performance of the three techniques we use the Arena-Hard-Auto benchmark. This benchmark is designed to be highly correlated with the performance of the models on the [LMSYS Chatbot Arena Leaderboard](). We use the newly released gpt-4o-mini as the base model. The results are shown below.

| Model | Score | 95% CI | Average #tokens |
|---|---|---|---|
| moa-gpt-4o-mini | 85.6 | (-1.7, 1.7) | 733 |
| gpt-4-turbo-2024-04-09 | 82.6 | (-1.6, 2.0) | 662 |
| claude-3.5-sonnet-20240620 | 79.3 | (-1.8, 2.1) | 567 |
| gpt-4o-2024-05-13 | 79.2 | (-1.8, 1.5) | 696 |
| gpt-4-0125-preview | 78.0 | (-1.6, 1.7) | 619 |
| bon-gpt-4o-mini | 75.0 | (-2.0, 2.3) | 659 |
| mcts-gpt-4o-mini | 74.8 | (-2.3, 1.9) | 663 |
| gpt-4o-mini | 74.1 | (-2.0, 2.0) | 670 |

We found that all the techniques bon, mcts and moa improve the performance when compared to the base model. In fact, with Patched MOA we were able to beat (by 3 points) even the gpt-4-turbo-2024-04-09 model which is currently the best model on the benchmark. When we compare the price of gpt-4-turbo with gpt-4o-mini we see that we are able to provide better performance at 1/50th the cost even when accounting for all the additional calls and tokens needed for Patched MOA.

## Patchflows

Next, we apply Patched MOA to compare the performance of different patchflows. The following table shows the numbers for each of the patchflows supported by our open-source framework patchwork. We selected a sample of the most active GitHub repositories in 3 different languages (Python, Java and JavaScript). Then we ran the patchflows on these repositories including their issues and pull requests on the main branch. We ran each patchflow only once; however a patchflow may make several calls to the LLM during the run depending on how it is implemented.

| Patchflow | Base RTC Eval | Optimized RTC Eval |
|---|---|---|
| AutoFix | 41.18 | 46.67 |
| PRReview | 50 | 100 |
| GenerateDocstring | 71.21 | 89.52 |
| GenerateREADME | 66.67 | 71.43 |
| ResolveIssue | 61.11 | 85.71 |

The Base RTC Eval shows the pass rate with the base model and the Optimized RTC Eval shows the pass rate with optimized inference using Patched MOA. We use Patched RTC as our evaluation metric as in prior work we showed that it is a good self evaluation metric that is correlated with accuracy on diverse downstream tasks. As is clear from the table above, we see better performance across all the patchflows. Using Patched MOA is an easy way to improve the accuracy of your patchflows without requiring any changes to prompts or the implementation of the development task.

# Discussion

Our research into inference optimization techniques for large language models (LLMs) has revealed several important insights and trade-offs that warrant further discussion. While we evaluated three distinct approaches—Best of N (bon), Monte Carlo Tree Search (mcts), and Mixture of Agents (moa)—our results led us to focus primarily on MOA for Patched MOA. This decision was based on a careful consideration of performance improvements, computational costs, and practical applicability across diverse software development tasks.

Our evaluation of three inference optimization techniques revealed distinct trade-offs between performance improvement and computational cost. Best of N (bon) offered modest gains (74.1 to 75.0 on the Arena-Hard-Auto benchmark) with minimal overhead (4x API calls, 3x time), while Monte Carlo Tree Search (mcts) showed similar improvement (74.8) but at a significantly higher computational cost (9x API calls, 8x time). In contrast, Mixture of Agents (moa) emerged as the superior approach, dramatically boosting performance from 74.1 to 85.6, surpassing even larger models like gpt-4-turbo, while maintaining a moderate computational overhead (3x API calls, 8x time). This analysis clearly demonstrates MOA's exceptional balance of performance enhancement and resource efficiency, justifying its selection as the core of our Patched MOA approach.

# Conclusions

In this work, we introduced Patched MOA, an inference optimization technique that beats GPT-4 over a wide range of tasks at 1/50th of the cost. Our technique is completely transparent to the user and can be applied to any LLM entirely during the inference. We also show that Patched MOA improves on Patched RTC [4] based evaluation for different patchflows that correspond to diverse software development tasks. Our technique is open-source and implemented in optillm an open-source optimizing inference proxy [5].

## Usage

To get access to Patched MOA:

Use the `patched_api_key` with our OpenAI compatible endpoint available at [patched.codes](patched.codes) and just use the base url `https://patchwork.patched.codes/optimize/v1`. If you want to compare with how the response would have been without Patched MOA, you can send the same request through our usual OpenAI compatible endpoint at `https://patchwork.patched.codes/v1`.